\documentclass[aps,twocolumn,preprintnumbers,amsmath,amssymb,floats,citeautoscript]{revtex4-1}
\usepackage{graphicx}
\usepackage{dcolumn}
\usepackage{bm}

\begin{document}

\title{Nodeless vs nodal order parameters in LiFeAs and LiFeP superconductors}

\author{K.~Hashimoto$^1$}
\author{S.~Kasahara$^2$}
\author{R.~Katsumata$^1$}
\author{Y.~Mizukami$^1$}
\author{M.~Yamashita$^1$}
\author{H.~Ikeda$^1$}
\author{T.~Terashima$^2$}
\author{A.~Carrington$^3$}
\author{Y.~Matsuda$^1$}
\author{T.~Shibauchi$^1$}

\affiliation{$^1$Department of Physics, Kyoto University, Sakyo-ku, Kyoto 606-8502, Japan\\
$^2$Research Center for Low Temperature and Materials Sciences, Kyoto University, Kyoto 606-8502, Japan\\
$^3$H. H. Wills Physics Laboratory, University of Bristol, Tyndall Avenue, Bristol, UK }

\date{\today}



\begin{abstract}
High-precision measurements of magnetic penetration depth $\lambda$ in clean single crystals of LiFeAs and LiFeP
superconductors reveal contrasting low-energy quasiparticle excitations. In LiFeAs the low-temperature $\lambda(T)$
shows a flat dependence indicative of a fully gapped state, which is consistent with previous studies. In contrast,
LiFeP exhibits a $T$-linear dependence of superfluid density $\propto \lambda^{-2}$, indicating a nodal superconducting
order parameter. A systematic comparison of quasiparticle excitations in the 1111, 122, and 111 families of
iron-pnictide superconductors implies that the nodal state is induced when the pnictogen height from the iron plane
decreases below a threshold value of $\sim 1.33$\,\AA.
\end{abstract}

\maketitle


There is growing evidence that the superconducting gap structure is not universal in the iron-based superconductors
\cite{Stewart11}. In certain materials such as optimally doped (Ba,K)Fe$_2$As$_2$ and Ba(Fe,Co)$_2$As$_2$, strong
evidence for the fully gapped superconducting state has been observed from several low-energy quasiparticle excitation
probes including magnetic penetration depth \cite{Hashimoto09a,Luan11} and thermal conductivity measurements
\cite{Luo09,Tanatar10}. In contrast, significant excitations at low temperatures due to nodes in the energy gap have
been detected in several Fe-pnictide superconductors. These include LaFePO ($T_c=6$\,K)
\cite{Fletcher09,Hicks09,Yamashita09}, BaFe$_2$(As,P)$_2$ ($T_c\le31$\,K) \cite{Hashimoto10a,Nakai10,Yamashita11}, and
KFe$_2$As$_2$ ($T_c=4$\,K) \cite{Fukazawa09,Dong10,Hashimoto10b}. It is quite extraordinary that such distinct pairing
states appear in closely related members of the same class of superconductors. To understand the mechanism of
superconductivity in iron-based superconductors, it is essential to identify what determines nodal and nodeless states
\cite{Mazin08,Kuroki09,Chubukov09,Graser09,Ikeda10,Thomale11,Kontani10}.

\begin{figure}[b]
\includegraphics[width=0.97\linewidth]{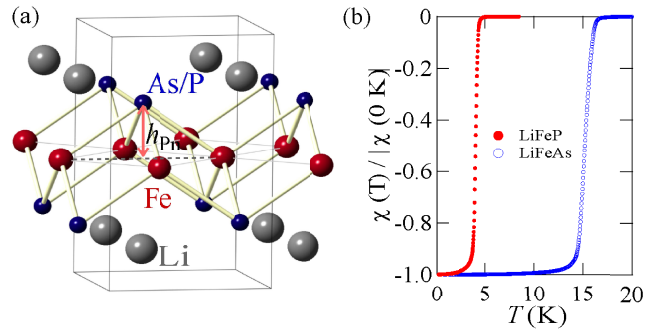}
\caption{ (Color online). (a) Schematic crystal structure of LiFe$Pn$ ($Pn=$ As or P). The arrow defines the pnictogen height $h_{Pn}$ from the iron plane. (b) The ac susceptibility of LiFe$Pn$ crystals measured from the frequency shift of the TDO.
} \label{transition}
\end{figure}

Theories based on antiferromagnetic spin fluctuations suggest that the pnictogen height $h_{Pn}$ above the iron plane
(see Fig.\:\ref{transition}(a)) is an important factor in determining the structure of the superconducting order
parameter \cite{Kuroki09,Graser09,Ikeda10,Thomale11}.  Generally, $h_{Pn}$ is much shorter for the P based
iron-pnictides in comparison to their As counterparts, so a good test of the theory would be to systematically compare
As and P based superconductors.  Although this can be achieved in part in the BaFe$_2$(As,P)$_2$ series, the fully As
containing end member BaFe$_2$As$_2$ is a nonsuperconducting antiferromagnet.  The same is true for LaFeAsO which
is the As analogue of the nodal superconductor LaFePO.  Charge doping of the arsenides induces superconductivity, but
also introduces disorder which complicates the identification of the pairing state.

The 111 materials, LiFeAs \cite{Wang08,Tapp08} and LiFeP \cite{Deng09,Mydeen10} provide a unique route to study this
problem as both materials are superconducting ($T_c\approx17$\,K and 4.5\,K, respectively), nonmagnetic, and importantly
very clean, with long electronic mean-free paths.  In LiFeAs, antiferromagnetic fluctuations have been observed
\cite{Jeglic11,Taylor11} and fully gapped superconductivity has been demonstrated by several experiments
\cite{Inosov10,Kim11,Imai11,Tanatar11,Li10,Borisenko10}, but no information has been reported for the pairing state in
LiFeP. Band-structure calculations show that the two materials exhibit similar Fermi surface shapes
\cite{Singh08,Shein10}; quasi-cylindrical hole sheets near the zone center and two warped electron sheets
near the zone corner.

Here we report on precision measurements of the magnetic penetration depth $\lambda(T)$ in single crystals, which
demonstrate a nodal gap state in LiFeP in sharp contrast to the nodeless state in LiFeAs. Our analysis based on
accumulated $\lambda(T)$ data in the 1111, 122, and 111 series of superconductors, indicates that the nodal state is
induced when $h_{Pn}$ is below a threshold value. By comparing calculated electronic band structures of LiFeAs and
LiFeP, we discuss the origin of this behavior.


Single crystals of LiFe$Pn$ ($Pn=$ As or P) were grown by a flux method \cite{Kasahara11}. The crystal size of LiFeP is
up to $135 \times 135 \times \sim10\,\mu$m, which is smaller than that of LiFeAs. To avoid degradation of the
sample due to reaction with air, the crystals were handled in an argon glove box and encapsulated in degassed Apiezon N
grease before measurements. Large residual resistivity ratios ($\sim50$ for LiFeAs and $\sim150$ for LiFeP)
\cite{Kasahara11}, observations of de Haas-van Alphen (dHvA) oscillations in magnetic torque \cite{Carrington11}, and
sharp superconducting transitions (Fig.\:\ref{transition}(b)) show that the crystals are of very high quality. The
temperature dependence of change in the magnetic penetration depth was measured by the tunnel diode oscillator (TDO)
technique \cite{Fletcher09,Hashimoto10b} down to $T/T_c\approx 0.03$. A weak ac field is applied along the $c$ axis so
that the supercurrent flows in the $ab$ plane.


\begin{figure}[tb]
\includegraphics[width=0.97\linewidth]{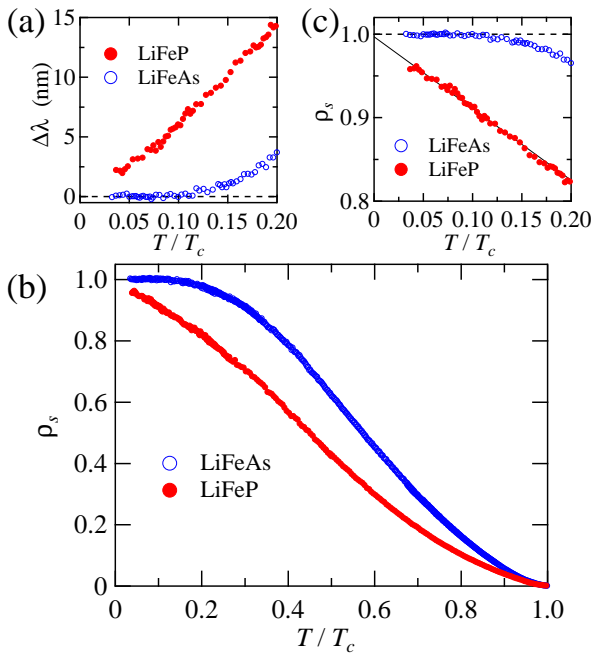}
\caption{ (Color online). (a) Low-temperature change in the magnetic penetration depth in single crystals of LiFeAs and LiFeP. (b) Temperature dependence of normalized superfluid density $\rho_s(T)$. We used $\lambda(0)=210$ and 150\,nm for LiFeAs and LiFeP, respectively. (c) Expanded view of $\rho_s(T)$ of LiFeP at low temperatures. The solid line is a fit to the $T$-linear dependence.
} \label{lambda}
\end{figure}

Figure\:\ref{lambda}(a) depicts the low-temperature variation of the in-plane penetration depth
$\Delta\lambda(T)=\lambda(T)-\lambda(0)$. The data for LiFeAs is completely flat within the experimental error of $\sim
0.3$\,nm below $T/T_c\sim 0.1$. This demonstrates negligible quasiparticle excitations at low temperatures, indicating
a fully gapped state. This result is fully consistent with previous results in LiFeAs \cite{Kim11,Imai11}. In sharp
contrast to this, the data for LiFeP exhibits much steeper temperature dependence of $\Delta\lambda(T)$ at low
temperatures. When we use a power law fit $\Delta\lambda(T)\propto T^\alpha$ to this data below $T/T_c\sim 0.25$, we
obtain a small value of $\alpha\approx 1.3$. In iron-based superconductors, a power law dependence with $\alpha\sim2$
can be expected even in the dirty full gap case when the sign changing $s_\pm$ state is considered \cite{Vorontsov09},
and indeed a tendency of the exponent decrease from $\alpha\gtrsim3$ to $\sim2$ with increased impurity scattering has
been observed experimentally \cite{Hashimoto09a,Kim10}. However, the small power $\alpha\lesssim1.5$ cannot be
explained by such a dirty nodeless state, and it is rather a strong indication that the superconducting gap has line
nodes. Indeed, our data can also be fitted to $\propto T^2/(T+T^*)$, which is applicable to the nodal case with small
impurity scattering \cite{Hirschfeld93}. The obtained low value of $T^*\approx0.3$\,K indicates a clean nodal behavior
and is consistent with the other measures of sample quality described above.

We also analyze the normalized superfluid density $\rho_s(T)=\lambda^2(0)/\lambda^2(T)$ (Fig.\:\ref{lambda}(b)). To do this we
need the value of $\lambda(0)$, which we cannot directly determine from the TDO measurements. The small angle neutron
scattering measurements of LiFeAs reveal $\lambda(0)\approx210$\,nm \cite{Inosov10}. To estimate the $\lambda(0)$ value
for LiFeP, we consider the difference of the effective mass in these two superconductors whose carrier number (Fermi
surface volume) is quite similar. The effective masses determined by the dHvA oscillations \cite{Carrington11} as well
as the electronic specific heat coefficients $\gamma$ have a factor of $\sim2$ difference ($\gamma\approx 16$ and
$\approx 30$\,mJ/K$^2$mol for LiFeP \cite{Deng09} and LiFeAs \cite{Lee10}, respectively), from which we estimate
$\lambda(0)\approx150$\,nm. The extracted temperature dependence of $\rho_s(T)$ shows contrasting behaviors for As and
P cases at low temperatures again: flat dependence for As and steeper dependence for P. The expanded view at low
temperatures (Fig.\:\ref{lambda}(c)) demonstrates a wide temperature range of $T$-linear dependence, which clearly
indicates the energy-linear density of state of quasiparticles and hence the existence of line nodes in the energy gap.


The strength of the electron-electron correlations can be measured by the mass enhancement which is closely related to
the $\gamma$ value. The larger $\gamma$ for LiFeAs than for LiFeP suggests weaker correlations in the P case, which is
reinforced by the smaller $A$ value of the Fermi-liquid coefficient in the $AT^2$ dependence of resistivity
\cite{Kasahara11,Heyer10} and smaller quasiparticle mass enhancements measured by quantum oscillations
\cite{Carrington11}. Strong correlations usually promote sign change in the superconducting order parameter \cite{Hashimoto10b}, which leads to the gap nodes in single-band superconductors. In the present multiband case with separated Fermi surface sheets, however, the seemingly opposite trend that LiFeP has nodes but is weakly correlated suggests that other factors are also important for node formation.

\begin{figure}[tb]
\includegraphics[width=0.97\linewidth]{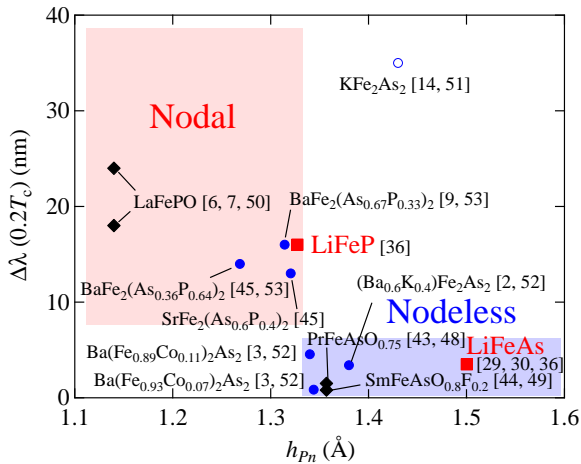}
\caption{ (Color online). Pnictogen-height dependence of $\Delta\lambda(T)$ at $T=0.2T_c$ as a measure of low-temperature quasiparticle excitations in 1111 (diamonds), 122 (circles), and 111 (squares) series of iron-pnictide superconductors. Here the pnictogen height $h_{Pn}$ (see Fig.\:\ref{transition}(a)) is determined from the structural analysis at room temperature \cite{Kasahara10,Kodama11,Zhigadlo08,Tegel08,Rotter08,Drotziger10,Kasahara11}. 
} \label{nodes}
\end{figure}

To obtain further insights, we gather the available data for the low-energy quasiparticle excitations in several
iron-pnictide superconductors including 1111 \cite{Fletcher09,Hicks09,Hashimoto09b,Malone09}, 122
\cite{Hashimoto10a,Hashimoto09a,Hashimoto10b,Hashimoto11,Luan11}, and 111-based materials \cite{Kim11,Imai11}. The
amount of thermally excited quasiparticles is directly related to the change in the penetration depth. Thus we quantify
$\Delta\lambda(0.2T_c)$ as a measure of excitations so that we avoid ambiguity resulting from uncertainties in
$\lambda(0)$. Among the available data in the literature, we select only the data which shows either $\alpha<1.5$ or
$>2.5$ in the power-law approximation, because the power-law dependence with $\alpha\sim 2$ cannot distinguish the
dirty nodeless and nodal states as discussed previously. A plot of $\Delta\lambda(0.2T_c)$ as a function of pnictogen
height $h_{Pn}$ in Fig.\:\ref{nodes} suggests that there is a threshold value of $\sim 1.33$\,\AA, below which all the
superconductors exhibit significant quasiparticle excitations (with $\alpha<1.5$) characteristic of a nodal state.
Above the threshold, most of the materials are nodeless with the exception of the highly hole-doped compound,
KFe$_2$As$_2$. This particular material is unusual in that it lacks electron sheets and thus there is no interband
nesting. In addition, the quasiparticle effective mass is strongly enhanced \cite{Hashimoto10b,Sato09}, and $T_c$ is
very low ($\sim 4$\,K) so superconductivity may have a different origin to that in the other materials. Therefore, our
analysis strongly suggests that the pnictogen height is an important parameter that determines the gap structure in the
iron-pnictide superconductors having significant interband scattering. 
One may also ask about the Fe-$Pn$-Fe bond angle, but the nodal LiFeP has a closer angle (108.6$^\circ$) to the perfect tetrahedron value of 109.47$^\circ$ than the nodeless LiFeAs (102.8$^\circ$), from which we do not find any simple correlation between the bond angle and gap structure in iron-pnictide superconductors. 

The importance of the pnictogen height $h_{Pn}$ on the superconducting order parameter in iron-pnictides has been
suggested in theoretical considerations based on the antiferromagnetic spin fluctuation mechanism
\cite{Kuroki09,Graser09,Ikeda10,Thomale11}. When $h_{Pn}$ is low, one of the hole bands with $d_{xy}$ orbital
character, which is located near the $(\pi,\pi)$ position in the unfolded Brillouin zone (BZ), tends to sink below the
Fermi level. The disappearance of this Fermi surface makes interband electron-hole scattering weaker and hence the
importance of the scattering between electron sheets relatively greater,  promoting a sign change of the
superconducting gap (and hence nodes) on the electron sheets.

\begin{figure}[tb]
\includegraphics[width=0.97\linewidth]{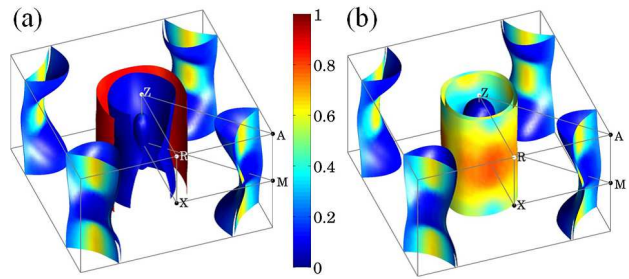}
\caption{ (Color online). Calculated Fermi surfaces of LiFeAs (a) and LiFeP (b). The spin-orbit interaction is included in the DFT calculations. 
Color indicates the relative weight of the $d_{xy}$ orbital contribution, which has been obtained from the Wannier fit by using \textsc{wannier90} \cite{WAN90} via \textsc{wien2wannier} interface \cite{W2W}.
} \label{bands}
\end{figure}

To check this, we have performed band-structure calculations based on density functional theory (DFT) including spin-orbit coupling, by using the
\textsc{wien2k} package \cite{WIEN2k} and the experimental lattice constants and internal positions \cite{Tapp08,Kasahara11}. The obtained Fermi surfaces  (Fig.\:\ref{bands}) are
similar to previous calculations \cite{Shein10}. Although angle-resolved photoemission results have suggested quite
different Fermi surfaces with no interband nesting in LiFeAs \cite{Borisenko10}, more recent bulk measurements of dHvA
oscillations reveal quasi-nested hole and electron sheets \cite{Carrington11} in a good agreement with the
calculations.  Importantly, the calculations show that the $d_{xy}$ hole sheet, which in the folded BZ is the outermost
hole sheet at $\Gamma$, is present in both compounds. This indicates that the absence of the $d_{xy}$ hole sheet is
not a requisite for the nodal state.

A more detailed comparison between LiFeAs and LiFeP reveals that the size of the outer hole sheet shrinks and its
relative weight of the $d_{xy}$ orbital contribution is significantly suppressed for the P case. Moreover, the middle
hole sheet in LiFeP has rather mixed $d_{xy}$ and $d_{zx/yz}$ contributions. Such differences in the orbital character
in hole sheets may affect relative importance of the interband hole-electron scattering compared with the scattering
between electron bands. dHvA measurements in LiFeP \cite{Carrington11} imply that the middle hole sheets has
 less mass enhancement than the electron sheets, suggesting that scattering between electron sheets is stronger
than the interband electron-hole scattering. This could lead to the formation of line nodes in the electron
sheets. We note that the extended-$s$ state with line nodes in the electron sheets has been discussed as the most
likely nodal gap structure of BaFe$_2$(As,P)$_2$ \cite{Yamashita11}. The strong $T$-linear dependence of superfluid
density in LiFeP is consistent with the nodes being on electron bands containing high Fermi velocity parts, which almost coincide with the $d_{xy}$-dominated regions (yellow parts of the electron sheets in Fig.\:\ref{bands}(b)) \cite{Kasahara11}. 
To determine the exact node locations in LiFeP, however, other measurements are necessary including angle-resolved probes of low-energy quasiparticle excitations. 

Our results that the nodal state is favored for low $h_{Pn}$ support the trend that the spin-fluctuation theory
predicts, but there remains challenging issues including the fact that the emergence of nodes is not directly caused by
the disappearance of the $d_{xy}$ hole sheet.
It has also been theoretically suggested that a competition between the orbital fluctuations and spin fluctuations
generates nodes in the electron sheets \cite{Kontani10}. 
The difference in the orbital character in hole sheets would also change the orbital fluctuations, which may affect the competition and hence the nodal gap structure. Further quantitative calculations of the pnictogen-height effect based on these theories
will help clarify the mechanism of iron-based superconductivity.


In summary, we have measured the penetration depth in clean crystals of LiFeAs and LiFeP. We found a $T$-linear superfluid density for
LiFeP indicating a nodal order parameter, in strong contrast to the fully-gapped state found for LiFeAs. A comparison of low-energy excitations across the different iron-pnictide superconductors suggests that the nodal state
is induced when the pnictogen height is shortened below a threshold value.

After completion of this study, we become aware of recent thermal conductivity results \cite{Qiu11} which suggest a
nodal state in Ba(Fe$_{0.64}$Ru$_{0.36}$)$_2$As$_2$ with $h_{Pn}\approx1.317$\,\AA \cite{Albenque10}, which supports our conclusion.


We thank K. Cho, A.\,V. Chubukov, A.\,I. Coldea, P.\,J. Hirschfeld, H. Kontani, K. Kuroki, I.\,I. Mazin, and R. Prozorov for discussions. This work is supported by KAKENHI from JSPS, Grant-in-Aid for GCOE
program ``The Next Generation of Physics, Spun from Universality and Emergence'' from MEXT, Japan, and EPSRC in the UK.


\end{document}